\documentclass{article}

\usepackage{arxiv}
\usepackage{bussproofs}
\usepackage{amsmath}
\usepackage{color}
\usepackage{listings}
\usepackage{amssymb}

\definecolor{dkgreen}{rgb}{0,0.6,0}
\definecolor{gray}{rgb}{0.5,0.5,0.5}
\definecolor{mauve}{rgb}{0.58,0,0.82}

\title{Tutorial implementation of Hoare logic in Haskell}

\author{
  Boro Sitnikovski \\
  Skopje, North Macedonia \\
  \texttt{buritomath@gmail.com} \\
}

\begin{document}
\maketitle

\begin{abstract}
Using the programming language Haskell, we introduce an implementation of propositional calculus, number theory, and a simple imperative language that can evaluate arithmetic and boolean expressions. Finally, we provide an implementation of Hoare's logic which will allow us to deduce facts about programs without the need for a full evaluation.
\end{abstract}


\section{Introduction}

Imperative programming languages run the world, to list a few popular ones: C, Python, JavaScript, PHP. Code is usually written in these languages by using an imperative style; that is, the computer is being told ("commanded") what to do and how to do it specifically. (The mathematical language is very unlike this, it is more declarative rather than imperative, that is, it doesn't care about the how.)

Since a programmer has to specify the exact "how", it is easy to make a programming error. In addition, given the popularity of these programming languages, there are many bugs in software applications that are programmed in them. This motivates for a way to formally verify certain properties about programs written in these languages. Hoare logic is one way to mechanically reason about computer programs.

We will first provide an implementation of systems for manipulating logic and numbers. Further, we will provide an implementation of an imperative language that will allow us to evaluate computer programs, and then, implementation of Hoare's logic, which will allow us to reason about computer programs, rather than evaluating them. The implementation of the imperative language will rely on logic for evaluating boolean expressions, while the implementation of Hoare's logic will rely on the number-theoretical system for proofs. Our implementation of Hoare's logic will be bottom-up, in the sense that we build the proofs from the ground up, in contrast to e.g. programming languages such as Dafny\cite{b1}, that build proofs from top to bottom. That is, in Dafny, the user provides a proposition and Dafny will derive the proof, automatically, using the automated theorem prover Z3\cite{b2}.

Hoare logic that we implement already has an implementation in Coq\cite{b3}. However, implementing a logic in Haskell is more concerned about playing at the value level (and to some extent at the type level), whereas in a dependently typed language the focus is at the type level. More specifically, in Haskell we cannot use any of the meta language's constructs to do mathematical proofs, so we have to take care of these algorithms ourselves.

Haskell is not a strongly normalizing language, which means that not every evaluation necessarily terminates. However, in dependently typed languages, such as Coq, the evaluation of proof terms (e.g. the type checker) is strongly normalizing, and this is what allows us to express mathematical proofs.

There are good introductory writings on dependent types\cite{b3}; there is a gentler approach that might be handy for newcomers\cite{b4}. There are good introductions to Haskell\cite{b5}, as well as mathematical logic\cite{b6}.

\section{Propositional calculus}

In this section, we provide an implementation of Propositional calculus, as described in GEB\cite{b7}.

\subsection{Syntax}

The syntax of the formal system expressed in BNF (Backus-Naur form) and in Haskell is:

\begin{minipage}{0.49\textwidth}
\begin{lstlisting}
prop   ::= bvar | prop brelop prop | unop prop
bvar   ::= A | B | C
unop   ::= "!"
brelop ::= "&&" | "||" | "->"
\end{lstlisting}
\end{minipage}
\begin{minipage}{0.49\textwidth}
\begin{lstlisting}
data PropCalc = A | B | C | Not PropCalc | And PropCalc PropCalc | Or PropCalc PropCalc | Imp PropCalc PropCalc
\end{lstlisting}
\end{minipage}

We will use the following, more generic implementation - instead of using specific variables, we rely on polymorphic types. This allows for an easy embedding of the system in other systems.

\begin{lstlisting}
data PropCalc a = PropVar a | Not (PropCalc a) | And (PropCalc a) (PropCalc a) | Or (PropCalc a) (PropCalc a) | Imp (PropCalc a) (PropCalc a)
\end{lstlisting}

We now have a way to represent some logical formulas, e.g. $A \land B$ as \texttt{And (PropVar A) (PropVar B)}.

We also need a way to differentiate between well-formed formulas and theorems since not all well-formed formulas are theorems. For that, we provide the \texttt{Proof} data type constructor:

\begin{lstlisting}
newtype Proof a = Proof a

fromProof :: Proof a -> a
fromProof (Proof a) = a
\end{lstlisting}

Note that \texttt{Proof \$ And (PropVar A) (PropVar B)} is different from \texttt{And (PropVar A) (PropVar B)}. However, the constructor mustn't be used directly; proofs should only be constructed given the rules that we provide next.

\subsection{Rules}

These rules come from GEB, and we list them here for completeness:

\begin{itemize}
\item \textbf{Fantasy Rule (Implication-introduction)}: If $x$ were a theorem, $y$ would be a theorem ($<x \to y>$).
\item \textbf{Carry Over Rule}: Inside a fantasy, any theorem from the "reality" a level higher can be brought in and used.
\item \textbf{Contrapositive Rule}: $< x \to y>$ and $< \neg y \to \neg x>$ are interchangeable.
\item \textbf{De Morgan's Rule}: $< \neg x \land \neg y>$ and $\neg<x \lor y>$ are interchangeable.
\item \textbf{Double-Tilde Rule (Double negation)}: The string $\neg\neg$ can be deleted from any theorem. It can also be inserted into any theorem, provided that the resulting string is itself well-formed.
\item \textbf{Joining Rule (And-introduction)}: If $x$ and $y$ are theorems, then $<x \land y>$ is a theorem.
\item \textbf{Rule of Detachment (Implication-elimination)}: If $x$ and $<x \to y>$ are both theorems, then $y$ is a theorem.
\item \textbf{Sep Rule (And-elimination)}: If $<x \land y>$ is a theorem, then both $x$ and $y$ are theorems.
\item \textbf{Switcheroo Rule}: $<x \lor y>$ and $<\neg x \to y>$ are interchangeable.
\end{itemize}

In Coq, each implementation of the rules would be represented at the type level. However, since we're working at the value level in Haskell, we rely on the \texttt{Either} data type to distinguish between provable and not provable terms; this abstraction increases complexity but allows for graceful error handling. We create a wrapper for that:

\begin{lstlisting}
type ESP a = Either String (Proof (PropCalc a))
\end{lstlisting}

Here's how we can implement \textbf{Joining Rule} and \textbf{Sep Rule}:

\begin{lstlisting}
ruleJoin :: Proof (PropCalc a) -> Proof (PropCalc a) -> ESP a
ruleJoin (Proof x) (Proof y) = Right $ Proof $ And x y
\end{lstlisting}

\begin{minipage}[t]{0.51\textwidth}
\begin{lstlisting}
ruleSepL :: Proof (PropCalc a) -> ESP a
ruleSepL (Proof (And x y)) = Right $ Proof x
ruleSepL _ = Left "ruleSepL: Cannot construct proof"
\end{lstlisting}
\end{minipage}
\begin{minipage}[t]{0.49\textwidth}
\begin{lstlisting}
ruleSepR :: Proof (PropCalc a) -> ESP a
ruleSepR (Proof (And x y)) = Right $ Proof y
ruleSepR _ = Left "ruleSepL: Cannot construct proof"
\end{lstlisting}
\end{minipage}

That is, we just lift the values from the object level to Haskell's level, and Haskell's neat type system takes care of everything else. Even though \texttt{ruleJoin} cannot fail (always returns a proof), it's still a good practice to return \texttt{ESP} for easy composition between rules.

Perhaps the most powerful rule is \textbf{Fantasy Rule}, and its counter-part \textbf{Rule of Detachment}:

\begin{lstlisting}
ruleFantasy :: PropCalc a -> (Proof (PropCalc a) -> ESP a) -> ESP a
ruleFantasy x f = f (Proof x) >>= \(Proof y) -> Right $ Proof $ Imp x y

ruleDetachment :: Eq a => Proof (PropCalc a) -> Proof (PropCalc a) -> ESP a
ruleDetachment (Proof x) (Proof (Imp x' y)) | x == x' = Right $ Proof y
ruleDetachment _ _ = Left "ruleDetachment: Cannot construct proof"
\end{lstlisting}

Note how we accept a non-proven term within \texttt{ruleFantasy}, whereas in other rules we accept proven terms; the hypothesis needn't be necessarily true, it only states that "If this hypothesis were a theorem, that would be a theorem". Further, we use the \texttt{Either} monad to cascade success or halt on failure. We also rely on Haskell's machinery for functions (lambda calculus) to take care of doing the actual work; we get support for \textbf{Carry Over Rule} out of the box.

\begin{lstlisting}
> putStrLn $ pr $ ruleFantasy (And (PropVar A) (PropVar B)) (\pq -> join $ ruleJoin <$> ruleSepR pq <*> ruleSepL pq)
⊢ A∧B→B∧A ✓
\end{lstlisting}

Here's another example implementation for \textbf{Double-Tilde Rule}:

\begin{minipage}{0.45\textwidth}
\begin{lstlisting}
ruleDoubleTildeIntro :: Proof (PropCalc a) -> ESP a
ruleDoubleTildeIntro (Proof x) = Right $ Proof $ Not (Not x)
\end{lstlisting}
\end{minipage}
\begin{minipage}{0.53\textwidth}
\begin{lstlisting}
ruleDoubleTildeElim :: Proof (PropCalc a) -> ESP a
ruleDoubleTildeElim (Proof (Not (Not x))) = Right $ Proof x
ruleDoubleTildeElim _ = Left "ruleDoubleTildeElim: Cannot construct proof"
\end{lstlisting}
\end{minipage}

However, in the current implementation, we can't apply a rule to e.g. $B$ within $A \lor B$, because it only accepts a full formula and doesn't know how to do replacements in its subformulas. To address this, we create a function that accepts a "path" (which subformula to apply it to), what rule should be applied, and the formula itself.

\begin{lstlisting}
data Pos = GoLeft | GoRight
type Path = [Pos]

applyPropRule :: Path -> (Proof (PropCalc a) -> ESP a) -> Proof (PropCalc a) -> ESP a
applyPropRule xs f (Proof x) = go xs f x
  where
  go :: Path -> (Proof (PropCalc a) -> ESP a) -> PropCalc a -> ESP a
  go (_:xs) f (Not x)         = go xs f x >>= \(Proof x) -> Right $ Proof $ Not x
  go (GoLeft:xs) f (And x y)  = go xs f x >>= \(Proof x) -> Right $ Proof $ And x y
  go (GoLeft:xs) f (Or x y)   = go xs f x >>= \(Proof x) -> Right $ Proof $ Or x y
  go (GoLeft:xs) f (Imp x y)  = go xs f x >>= \(Proof x) -> Right $ Proof $ Imp x y
  go (GoRight:xs) f (And x y) = go xs f y >>= \(Proof y) -> Right $ Proof $ And x y
  go (GoRight:xs) f (Or x y)  = go xs f y >>= \(Proof y) -> Right $ Proof $ Or x y
  go (GoRight:xs) f (Imp x y) = go xs f y >>= \(Proof y) -> Right $ Proof $ Imp x y
  go _ f x = f (Proof x)
\end{lstlisting}

Of course, going left or right doesn't make much sense for unary operators, so the implementation will just drill down in the \texttt{Not} operator if it finds either a left or a right. But, going left or right does make sense for binary operators - we drill on the left and the right argument respectively.

Now we can use this to apply \textbf{Double-Tilde Rule} to $B$ within $A \lor B$ as follows:

\begin{lstlisting}
> putStrLn $ pr $ ruleFantasy (Or (PropVar A) (PropVar B)) (applyPropRule [GoRight] ruleDoubleTildeIntro)
⊢ A∨B→A∨¬¬B ✓
\end{lstlisting}

However, consider the following:

\begin{lstlisting}
> putStrLn $ pr $ ruleFantasy (And (PropVar A) (PropVar B)) return >>= applyPropRule [GoLeft] ruleSepL
⊢ A→A∧B ✓
\end{lstlisting}

This utility function should be used with care, since we may end up with incorrect formulas. It is possible to derive the same formulas without using it at all, however in some cases using it adds convenience, at a price.

\begin{lstlisting}
> :{
| putStrLn $ pr $ ruleFantasy (Or (PropVar A) (PropVar B)) $ \premise -> do
|   step1 <- ruleSwitcheroo premise
|   step2 <- ruleContra step1
|   step3 <- ruleFantasy (Not (PropVar B)) $ \premise2 -> do
|     step4 <- ruleDetachment premise2 step2
|     ruleDoubleTildeElim step4
|   prfBorA <- ruleSwitcheroo step3
|   step5 <- ruleSwitcheroo prfBorA
|   step6 <- ruleContra step5
|   ruleSwitcheroo step6
| :}
⊢ A∨B→A∨¬¬B ✓
\end{lstlisting}

\section{Number theory}

In this section we provide an implementation of a number-theoretical system, using the same literature (GEB).

\subsection{Syntax}

We start by introducing a language that can represent arithmetic expressions.

\begin{minipage}{0.49\textwidth}
\begin{lstlisting}
digit  ::= "0" | "1" | ... | "8" | "9"
arith  ::= number | var | aterm relop aterm
number ::= digit*
relop  ::= "+" | "*"
var    ::= A | B | C ... | Y | Z
\end{lstlisting}
\end{minipage}
\begin{minipage}{0.49\textwidth}
\begin{lstlisting}
data Arith =
  Num Integer
  | Var Char
  | Plus Arith Arith
  | Mult Arith Arith
\end{lstlisting}
\end{minipage}

As we did with \texttt{PropCalc} earlier, we will use a more generic variant; getting rid of integers (implement our own zero and successor), and abstracting \texttt{Char}.

\begin{lstlisting}
data Arith a =
  Var a
  | Z
  | S (Arith a)
  | Plus (Arith a) (Arith a)
  | Mult (Arith a) (Arith a)
\end{lstlisting}

For example, we can now represent $0 + 1 \cdot A$ as \texttt{Plus Z (Mult (S Z) (Var A))}.

The next step is to implement the components from first-order logic (quantifiers and arithmetic equations).

\begin{minipage}{0.49\textwidth}
\begin{lstlisting}
fol    ::= eq arith arith | forall var prop | exists a prop
var    ::= A | B | C ... | Y | Z
\end{lstlisting}
\end{minipage}
\begin{minipage}{0.49\textwidth}
\begin{lstlisting}
data FOL a =
  Eq (Arith a) (Arith a)
  | ForAll a (PropCalc (FOL a))
  | Exists a (PropCalc (FOL a))
\end{lstlisting}
\end{minipage}

Note how we embed \texttt{PropCalc} within the system. Similarly, as before, these constructors must not be used directly; proofs should only be constructed given the rules, which we provide next.

\subsection{Rules}

These formation rules come from GEB as well:

\begin{itemize}
\item \textbf{Rule of Specification}: Suppose $u$ is a variable that occurs inside the string $x$. If the string $\forall u:x$ is a theorem, then so is $x$, and so are any strings made from $x$ by replacing $u$, wherever it occurs, by the same term. (Restriction: The term that replaces $u$ must not contain any variable quantified in $x$.)
\item \textbf{Rule of Generalization}: Suppose $x$ is a theorem in which $u$, a variable, occurs free. Then $\forall u:x$ is a theorem. (Restriction: No generalization is allowed in a fantasy on any variable which appeared free in the fantasy's premise.)
\item \textbf{Rule of Interchange}: Suppose $u$ is a variable. Then the strings $\forall u:\neg$ and $\neg \exists u:$ are interchangeable anywhere inside any theorem.
\item \textbf{Rule of Existence}: Suppose a term (which may contain variables as long as they are free) appears once, or multiply, in a theorem. Then any (or several, or all) of the appearances of the term may be replaced by a variable that otherwise does not occur in the theorem, and the corresponding existential quantifier must be placed in front.
\item \textbf{Rule of Symmetry}: If $r=s$ is a theorem, then so is $s=r$.
\item \textbf{Rule of Transitivity}: If $r=s$ and $s=t$ are theorems, then so is $r=t$.
\item \textbf{Rule Add S}: If $r=t$ is a theorem, then $Sr=St$ is a theorem.
\item \textbf{Rule Drop S}: If $Sr=St$ is theorem, then $r=t$ is a theorem.
\item \textbf{Rule of Induction}: Let $X[u]$ represent a well-formed formula in which the variable $u$ is free, and $X[x/u]$ represents the same string, with each appearance of $u$ replaced by $x$. If both $\forall u:X[u] \to X[Su/u]$ and $X[0/u]$ are theorems, then $\forall u:X[u]$ is also a theorem.
\end{itemize}

We provide an implementation of a function which replaces all occurences of a term with another term in a formula:

\begin{lstlisting}
substPropCalc :: Eq a => Proof (PropCalc (FOL a)) -> Arith a -> Arith a -> Proof (PropCalc (FOL a))
substPropCalc (Proof f) v e = Proof $ go f v e where
  go :: Eq a => PropCalc (FOL a) -> Arith a -> Arith a -> PropCalc (FOL a)
  go (PropVar (Eq a b)) v e     = PropVar (Eq (substArith a v e) (substArith b v e))
  go (PropVar (ForAll x y)) v e = PropVar (ForAll x (go y v e))
  go (PropVar (Exists x y)) v e = PropVar (Exists x (go y v e))
  go (Not x) v e                = Not (go x v e)
  go (And x y) v e              = And (go x v e) (go y v e)
  go (Or x y) v e               = Or (go x v e) (go y v e)
  go (Imp x y) v e              = Imp (go x v e) (go y v e)
\end{lstlisting}

The function \texttt{substArith} is similar to \texttt{substPropCalc}, defined for its own data type. Next, we provide the functions \texttt{getArithVars} and \texttt{getBoundVars} that retrieve all variables in an arithmetic expression and all bound variables in a formula respectively.

\begin{minipage}{0.49\textwidth}
\begin{lstlisting}
getArithVars :: Eq a => Arith a -> [a]
getArithVars x = nub $ go x where
  go (Var a) = [a]
  go (S x) = go x
  go (Plus a b) = go a ++ go b
  go (Mult a b) = go a ++ go b
  go _ = []
\end{lstlisting}
\end{minipage}
\begin{minipage}{0.49\textwidth}
\begin{lstlisting}
getBoundVars :: Eq a => PropCalc (FOL a) -> [a]
getBoundVars x = nub $ go x where
  go (PropVar (ForAll x y)) = x : go y
  go (PropVar (Exists x y)) = x : go y
  go _ = []
\end{lstlisting}
\end{minipage}

We are now able to implement \textbf{Rule of Specification}:

\begin{lstlisting}
ruleSpec :: Eq a => Arith a -> Proof (PropCalc (FOL a)) -> ESP (FOL a)
ruleSpec e (Proof (PropVar (ForAll x y))) | not (any (`elem` getArithVars e) (getBoundVars y)) = Right $ substPropCalc (Proof y) (Var x) e
ruleSpec _ _ = Left "ruleSpec: Cannot construct proof"
\end{lstlisting}

We can also implement \textbf{Rule of Generalization} as follows:

\begin{lstlisting}
ruleGeneralize :: Eq a => a -> [Proof (PropCalc (FOL a))] -> Proof (PropCalc (FOL a)) -> ESP (FOL a)
ruleGeneralize x premises (Proof y) | x `notElem` getBoundVars y && x `notElem` concatMap (getFreeVars . fromProof) premises -- fantasy vars
  = Right $ Proof $ PropVar (ForAll x y)
ruleGeneralize _ _ _ = Left "ruleGeneralize: Cannot construct proof"
\end{lstlisting}

The third argument represents the premises (if any) within fantasies - this is the Restriction part of the rule.

Another interesting rule is \textbf{Rule of Existence}. We can't simply substitute everything as we did with \texttt{ruleSpec}; we need to allow replacement of zero, one, or multiple terms, provided these terms are equivalent.

We're given the following:

\begin{itemize}
\item \texttt{applyFOLRule} and \texttt{applyArithRule}, both of which are similar to \texttt{applyPropRule}. (Restriction: \texttt{applyFOLRule} cannot be applied to any quantified (bound) variables, similarly to \textbf{Rule of Generalization}.)
\item \texttt{applyFOLArithRule} which is a combination of \texttt{applyFOLRule} and \texttt{applyArithRule}, which takes a list of paths and then applies the rule to all of them.
\item \texttt{getTerm} - similar to \texttt{applyArithRule}, except that it returns a term instead of applying any function to it. This function is used to check that we're replacing the one and the same term in the formula.
\end{itemize}

Now, \textbf{Rule of Existence} can be implemented as follows:

\begin{lstlisting}
ruleExistence :: Eq a => a -> [(Pos, Path, Path)] -> Proof (PropCalc (FOL a)) -> ESP (FOL a)
ruleExistence x paths (Proof y) | allSame (map (\path -> getTerm path y) paths) =
  go (Proof y) paths >>= \(Proof y) -> Right $ Proof $ PropVar (Exists x y)
  where
  go f ((pos, path1, path2):paths) = applyFOLArithRule pos path1 path2 (\_ -> Var x) f >>= \prf -> go prf paths
  go x _ = Right x
ruleExistence x [] (Proof y) | x `notElem` getBoundVars y =
  Right $ Proof $ PropVar (Exists x y)
ruleExistence _ _ _ = Left "ruleExistence: Cannot construct proof"
\end{lstlisting}

Here's another example implementation for \textbf{Rule of Induction}:

\begin{lstlisting}
ruleInduction :: Eq a => Proof (PropCalc (FOL a)) -> Proof (PropCalc (FOL a)) -> ESP (FOL a)
ruleInduction base (Proof ih@(PropVar (ForAll x (Imp y z)))) =
  -- in base' and conc, y is Proof y because it's an assumption
  let base' = substPropCalc (Proof y) (Var x) Z
      conc  = substPropCalc (Proof y) (Var x) (S (Var x)) in
  -- similarly, z is Proof z here
  if base' == base && conc == Proof z
  then Right $ Proof $ PropVar (ForAll x y)
  else Left "ruleInduction: Cannot construct proof"
ruleInduction _ _ = Left "ruleInduction: Cannot construct proof"
\end{lstlisting}

\subsection{Peano's axioms}

Finally, we now encode Peano's axioms as follows:

\begin{lstlisting}
axiom1 (Var a) = Right $ Proof $ PropVar (ForAll a (Not (PropVar (Eq (S (Var a)) Z))))
axiom1 _ = Left "axiom1: Cannot construct proof"

axiom2 (Var a) = Right $ Proof $ PropVar (ForAll a (PropVar (Eq (Plus (Var a) Z) (Var a))))
axiom2 _ = Left "axiom2: Cannot construct proof"

axiom3 (Var a) (Var b) = Right $ Proof $ PropVar (ForAll a (PropVar (ForAll b (PropVar (Eq (Plus (Var a) (S (Var b))) (S (Plus (Var a) (Var b))))))))
axiom3 _ _ = Left "axiom3: Cannot construct proof"

axiom4 (Var a) = Right $ Proof $ PropVar (ForAll a (PropVar (Eq (Mult (Var a) Z) Z)))
axiom4 _ = Left "axiom4: Cannot construct proof"

axiom5 (Var a) (Var b) = Right $ Proof $ PropVar (ForAll a (PropVar (ForAll b (PropVar (Eq (Mult (Var a) (S (Var b))) (Plus (Mult (Var a) (Var b)) (Var a)))))))
axiom5 _ _ = Left "axiom5: Cannot construct proof"
\end{lstlisting}

With a few example proofs, where we use the \texttt{do} notation to increase readability by avoiding nested monadic binds.

\begin{minipage}[t]{0.53\textwidth}
$\begin{aligned}
\star \ \forall C: \forall D: D + S(C) = S(D) + C \to \\
\forall D: D + S(S(C)) = S(D) + S(C):
\end{aligned}$

\begin{lstlisting}
lemma1 = do
  -- ⊢ ∀A:∀B:(A+SB=S(A+B))
  step1 <- axiom3 (Var A) (Var B)
  -- ⊢ ∀B:(D+SB=S(D+B))
  step2 <- ruleSpec (Var D) step1
  -- ⊢ D+SSC=S(D+SC)
  step3 <- ruleSpec (S (Var C)) step2
  -- ⊢ ∀B:(SD+SB=S(SD+B))
  step4 <- ruleSpec (S (Var D)) step1
  -- ⊢ SD+SC=S(SD+C)
  step5 <- ruleSpec (Var C) step4
  -- ⊢ S(SD+C)=SD+SC
  step6 <- ruleSymmetry step5
  -- ⊢ ∀D:(D+SC=SD+C)→∀D:(D+SSC=SD+SC)
  step7 <- ruleFantasy (PropVar (ForAll D (PropVar (Eq (Plus (Var D) (S (Var C))) (Plus (S (Var D)) (Var C)))))) $ \premise -> do
    -- ⊢ D+SC=SD+C
    step8 <- ruleSpec (Var D) premise
    -- ⊢ S(D+SC)=S(SD+C)
    step9 <- ruleAddS step8
    -- ⊢ D+SSC=S(SD+C)
    step10 <- ruleTransitivity step3 step9
    -- ⊢ D+SSC=SD+SC
    step11 <- ruleTransitivity step10 step6
    -- ⊢ ∀D:(D+SSC=SD+SC)
    ruleGeneralize D [premise] step11
  -- lemma1 ⊢ ∀C:<∀D:(D+SC=SD+C)→∀D:(D+SSC=SD+SC)>
  ruleGeneralize C [] step7
\end{lstlisting}
\end{minipage}
\begin{minipage}[t]{0.45\textwidth}
$\star \ \forall D: D + S(Z) = S(D) + Z$:

\begin{lstlisting}
lemma2 = do
  -- ⊢ ∀A:∀B:(A+SB=S(A+B))
  step1 <- axiom3 (Var A) (Var B)
  -- ⊢ ∀B:(D+SB=S(D+B))
  step2 <- ruleSpec (Var D) step1
  -- ⊢ D+SSC=S(D+SC)
  step3 <- ruleSpec (S (Var C)) step2
  -- ⊢ D+S0=S(D+0)
  step4 <- ruleSpec Z step2
  -- ⊢ ∀A:(A+0=A)
  step5 <- axiom2 (Var A)
  -- ⊢ D+0=D
  step6 <- ruleSpec (Var D) step5
  -- ⊢ S(D+0)=SD
  step7 <- ruleAddS step6
  -- ⊢ D+S0=SD
  step8 <- ruleTransitivity step4 step7
  -- ⊢ SD+0=SD
  step9 <- ruleSpec (S (Var D)) step5
  -- ⊢ SD=SD+0
  step10 <- ruleSymmetry step9
  -- ⊢ D+S0=SD+0
  step11 <- ruleTransitivity step8 step10
  -- lemma2 ⊢ ∀D:(D+S0=SD+0)
  ruleGeneralize D [] step11
\end{lstlisting}

$\star \ \forall C: \forall D: D + S(C) = S(D) + C$:

\begin{lstlisting}
-- ⊢ ∀C:∀D:(D+SC=SD+C)
theorem = join $ ruleInduction <$> lemma2 <*> lemma1
\end{lstlisting}
\end{minipage}

\section{Imperative language}

\subsection{Arithmetic and boolean expressions}

To start with, we add support for variables by introducing the notion of a context; one implementation in Haskell is simply a mapping from characters to numbers.

\begin{minipage}{0.49\textwidth}
\begin{lstlisting}
type Context = M.Map Char Integer
\end{lstlisting}
\end{minipage}
\begin{minipage}{0.49\textwidth}
\begin{lstlisting}
type Context a = M.Map a Integer
\end{lstlisting}
\end{minipage}

However, as with previous implementations, we keep things more generic and use the implementation on the right.

We list the evaluation rules for arithmetic expressions using natural deduction style; the expression above the line represents hypothesis, and the expression below represents conclusion:

\[
\AxiomC{$(\#v, \$v') \in ctx$}
\RightLabel{(A-Eval-Var)}
\UnaryInfC{$\#v \underset{ctx}{\to} \$v'$}
\DisplayProof
\quad
\AxiomC{}
\RightLabel{(A-Eval-Z)}
\UnaryInfC{$\texttt{Z} \underset{ctx}{\to} 0$}
\DisplayProof
\quad
\AxiomC{$a \underset{ctx}{\to} a'$}
\RightLabel{(A-Eval-Succ)}
\UnaryInfC{$\texttt{S} \ a \underset{ctx}{\to} 1 + a_1'$}
\DisplayProof
\]
\hfill
\[
\AxiomC{$a_1 \underset{ctx}{\to} a_1' , a_2 \underset{ctx}{\to} a_2'$}
\RightLabel{(A-Eval-Plus)}
\UnaryInfC{$\texttt{Plus} \ a_1 \ a_2 \underset{ctx}{\to} a_1' + a_2'$}
\DisplayProof
\quad
\quad
\AxiomC{$a_1 \underset{ctx}{\to} a_1' , a_2 \underset{ctx}{\to} a_2'$}
\RightLabel{(A-Eval-Mult)}
\UnaryInfC{$\texttt{Mult} \ a_1 \ a_2 \underset{ctx}{\to} a_1' \cdot a_2'$}
\DisplayProof
\]

The arrow symbol $\underset{ctx}{\to}$ in the rules represents the actual evaluation of arithmetic expressions under context $ctx$.

\begin{lstlisting}
aeval :: (Ord a, Eq a) => Context a -> Arith a -> Either String Integer
aeval ctx (Var v)        = if M.member v ctx then Right (ctx M.! v) else Left "Element not found"
aeval ctx Z              = Right 0
aeval ctx (S a)          = aeval ctx a >>= \a -> Right $ 1 + a
aeval ctx (Plus a1 a2)   = aeval ctx a1 >>= \a1 -> aeval ctx a2 >>= \a2 -> Right $ a1 + a2
aeval ctx (Mult a1 a2)   = aeval ctx a1 >>= \a1 -> aeval ctx a2 >>= \a2 -> Right $ a1 * a2
\end{lstlisting}

Further, we provide the evaluation rules for boolean expressions:

\[
\AxiomC{$a_1 \underset{ctx}{\to} a_1' , a_2 \underset{ctx}{\to} a_2'$}
\RightLabel{(B-Eval-Eq)}
\UnaryInfC{$\texttt{Eq} \ a_1 \ a_2 \underset{ctx}{\Rightarrow} a_1' = a_2'$}
\DisplayProof
\AxiomC{$b \underset{ctx}{\Rightarrow} b'$}
\RightLabel{(B-Eval-ForAll)}
\UnaryInfC{$\texttt{ForAll} \ x \ b \underset{ctx}{\Rightarrow} b'$}
\DisplayProof
\AxiomC{$b \underset{ctx}{\Rightarrow} b'$}
\RightLabel{(B-Eval-Exists)}
\UnaryInfC{$\texttt{Exists} \ x \ b \underset{ctx}{\Rightarrow} b'$}
\DisplayProof
\]
\hfill
\[
\AxiomC{$b \underset{ctx}{\Rightarrow} b'$}
\RightLabel{(B-Eval-Not)}
\UnaryInfC{$\texttt{Not} \ b' \underset{ctx}{\Rightarrow} \neg b'$}
\DisplayProof
\quad
\AxiomC{$b_1 \underset{ctx}{\Rightarrow} b_1' , b_2 \underset{ctx}{\Rightarrow} b_2'$}
\RightLabel{(B-Eval-And)}
\UnaryInfC{$\texttt{And} \ b_1 \ b_2 \underset{ctx}{\Rightarrow} b_1' \land b_2'$}
\DisplayProof
\AxiomC{$b_1 \underset{ctx}{\Rightarrow} b_1' , b_2 \underset{ctx}{\Rightarrow} b_2'$}
\RightLabel{(B-Eval-Or)}
\UnaryInfC{$\texttt{Or} \ b_1 \ b_2 \underset{ctx}{\Rightarrow} b_1' \lor b_2'$}
\DisplayProof
\]
\hfill
\[
\AxiomC{$b_1 \underset{ctx}{\Rightarrow} b_1' , b_2 \underset{ctx}{\Rightarrow} b_2'$}
\RightLabel{(B-Eval-Imp)}
\UnaryInfC{$\texttt{Imp} \ b_1 \ b_2 \underset{ctx}{\Rightarrow} \neg b_1' \lor b_2'$}
\DisplayProof
\]

Similarly, the double right arrow symbol $\underset{ctx}{\Rightarrow}$ in the rules represents the actual evaluation of boolean expressions under context $ctx$.

\begin{lstlisting}
beval :: (Ord a, Eq a) => Context a -> PropCalc (FOL a) -> Either String Bool
beval ctx (PropVar (Eq a1 a2))   = aeval ctx a1 >>= \a1 -> aeval ctx a2 >>= \a2 -> Right $ a1 == a2
beval ctx (PropVar (ForAll x b)) = beval ctx b
beval ctx (PropVar (Exists x b)) = beval ctx b
beval ctx (Not b)     = beval ctx b >>= \b -> Right $ not b
beval ctx (And b1 b2) = beval ctx b1 >>= \b1 -> beval ctx b2 >>= \b2 -> Right $ b1 && b2
beval ctx (Or b1 b2)  = beval ctx b1 >>= \b1 -> beval ctx b2 >>= \b2 -> Right $ b1 || b2
beval ctx (Imp b1 b2) = beval ctx b1 >>= \b1 -> beval ctx b2 >>= \b2 -> Right $ not b1 || b2
\end{lstlisting}

Note that we ignore quantifiers in this implementation because we only rely on propositional logic and substitute variables based on the context.

\begin{lstlisting}
> let e = Plus (Var A) (S Z) in putStrLn $ pr e ++ " == " ++ pr (aeval (M.fromList [(A, 5)]) e)
A+S0 == 6 ✓
> let e = Plus (Var A) (S (S Z)) in putStrLn $ pr e ++ " == " ++ pr (aeval (M.fromList [(A, 5)]) e)
A+SS0 == 7 ✓
> let e = PropVar $ Eq (Var A) (S Z) in putStrLn $ pr e ++ " == " ++ pr (beval (M.fromList [(A, 5)]) e)
A=S0 == False ✓
> let e = PropVar $ Eq (Var A) (S (S (S (S (S Z))))) in putStrLn $ pr e ++ " == " ++ pr (beval (M.fromList [(A, 5)]) e)
A=SSSSS0 == True ✓
\end{lstlisting}

\subsection{Commands}

We proceed with providing the syntax, evaluation rules, and their implementation in Haskell.

\begin{lstlisting}
data Command a =
  CSkip | CAssign a (Arith a) | CSequence (Command a) (Command a)
  | CIfElse (PropCalc (FOL a)) (Command a) (Command a) | CWhile (PropCalc (FOL a)) (Command a)
  | CAssert (PropCalc (FOL a)) (Command a) (PropCalc (FOL a))
\end{lstlisting}

That is, the language contains the minimum set of commands which make an imperative language:

\begin{itemize}
\item \texttt{CSkip} is the the no operation command - the empty statement.
\item \texttt{CAssign} will assign a value to a variable in a context.
\item \texttt{CSequence} will join two commands, which allows for the evaluation of commands in sequence.
\item \texttt{CIfElse} accepts a boolean and depending on its value either executes one command, or another.
\item \texttt{CWhile} accepts a boolean and keeps executing a command as long as the boolean is true.
\item \texttt{CAssert} accepts a precondition, a command, and a postcondition. The evaluation will be successful if the precondition and the postcondition are satisfied before and after executing the command, respectively.
\end{itemize}

We show the evaluation rules for this language:

\[
\AxiomC{}
\RightLabel{(C-Eval-Skip)}
\UnaryInfC{$\texttt{CSkip} \underset{ctx}{\mapsto} ctx$}
\DisplayProof
\]
\hfill
\[
\AxiomC{$v \underset{ctx}{\to} v'$}
\RightLabel{(C-Eval-Assign)}
\UnaryInfC{$\texttt{CAssign} \ c \ v \underset{ctx}{\mapsto} ctx \cup (c, v')$}
\DisplayProof
\quad
\AxiomC{$c_1 \underset{ctx}{\mapsto} ctx', c_2 \underset{ctx'}{\mapsto} ctx''$}
\RightLabel{(C-Eval-Sequence)}
\UnaryInfC{$\texttt{CSequence} \ c_1 \ c_2 \underset{ctx}{\mapsto} ctx''$}
\DisplayProof
\]
\hfill
\[
\AxiomC{$b \underset{ctx}{\Rightarrow} \texttt{True}, c_1 \underset{ctx}{\mapsto} ctx'$}
\RightLabel{(C-Eval-IfTrue)}
\UnaryInfC{$\texttt{CIfElse} \ b \ c_1 \ c_2 \underset{ctx}{\mapsto} ctx'$}
\DisplayProof
\quad
\AxiomC{$b \underset{ctx}{\Rightarrow} \texttt{False}, c_2 \underset{ctx}{\mapsto} ctx'$}
\RightLabel{(C-Eval-IfFalse)}
\UnaryInfC{$\texttt{CIfElse} \ b \ c_1 \ c_2 \underset{ctx}{\mapsto} ctx'$}
\DisplayProof
\]
\hfill
\[
\AxiomC{$b \underset{ctx}{\Rightarrow} \texttt{True}, c \underset{ctx}{\mapsto} ctx', \texttt{CWhile} \ b \ c \underset{ctx'}{\mapsto} ctx''$}
\RightLabel{(C-Eval-WhileTrue)}
\UnaryInfC{$\texttt{CWhile} \ b \ c \underset{ctx}{\mapsto} ctx''$}
\DisplayProof
\quad
\AxiomC{$b \underset{ctx}{\Rightarrow} \texttt{False}$}
\RightLabel{(C-Eval-WhileFalse)}
\UnaryInfC{$\texttt{CWhile} \ b \ c \underset{ctx}{\mapsto} ctx$}
\DisplayProof
\]
\hfill
\[
\AxiomC{$b_1 \underset{ctx}{\Rightarrow} \texttt{True}, c \underset{ctx}{\mapsto} ctx', b_2 \underset{ctx'}{\Rightarrow} \texttt{True}$}
\RightLabel{(C-Eval-Assert)}
\UnaryInfC{$\texttt{CAssert} \ b_1 \ c \ b_2$}
\DisplayProof
\]

The map arrow symbol $\underset{ctx}{\mapsto}$ in the rules represents the actual evaluation of a command under context $ctx$.

\begin{lstlisting}
eval :: (Ord a, Eq a) => Context a -> Command a -> Either String (Context a)
eval ctx CSkip             = Right ctx
eval ctx (CAssign c v)     = aeval ctx v >>= \v -> Right $ M.insert c v ctx
eval ctx (CSequence c1 c2) = let ctx' = eval ctx c1 in ctx' >>= (\ctx'' -> eval ctx'' c2)
eval ctx (CIfElse b c1 c2) = beval ctx b >>= \b -> eval ctx $ if b then c1 else c2
eval ctx (CWhile b c)      = beval ctx b >>= \b' ->
  if b'
  then let ctx' = eval ctx c in ctx' >>= (\ctx'' -> eval ctx'' (CWhile b c))
  else Right ctx
eval ctx (CAssert b1 c b2) = beval ctx b1 >>= \b1 ->
  if b1
  then eval ctx c >>=
       (\ctx' -> beval ctx' b2 >>= \b2 ->
                  if b2
                  then Right ctx'
                  else Left "Assert: Post-condition does not match!")
  else Left "Assert: Pre-condition does not match!"
\end{lstlisting}

Note that this language is not strongly normalizing; consider the evaluation of \texttt{CWhile BTrue CSkip}.

As an example, a program that counts from 0 to $B$ can be represented as follows:

\begin{minipage}{0.29\textwidth}
\begin{lstlisting}
A := 0
while ¬(A = B)
  A := 1 + A
\end{lstlisting}
\end{minipage}
\begin{minipage}{0.69\textwidth}
\begin{lstlisting}
countToB =
  let l1 = CAssign A Z
      l2 = CWhile (Not (PropVar $ Eq (Var A) (Var B))) l3
      l3 = CAssign A (S (Var A))
  in CSequence l1 l2
\end{lstlisting}
\end{minipage}

\pagebreak

Here's an example evaluation with assertion checks:

\begin{lstlisting}
> eval (M.fromList [(B, 3)]) countToB
Right (fromList [(A,3),(B,3)])
> let toArith x = if x == 0 then Z else S $ toArith (x - 1)
> let e = CAssert (PropVar $ Eq (toArith 5) (Var B)) countToB (PropVar $ Eq (Var B) (Var A)) in putStrLn $ "Assert {B=5} countToB {A=B}: " ++ show (eval (M.fromList [(B, 5)]) e)
Assert {B=5} countToB {A=B}: Right (fromList [(A,5),(B,5)])
> let e = CAssert (PropVar $ Eq (toArith 4) (Var B)) countToB (PropVar $ Eq (toArith 5) (Var A)) in putStrLn $ "Assert {B=4} countToB {A=5}: " ++ show (eval (M.fromList [(B, 5)]) e)
Assert {B=4} countToB {A=5}: Left "Assert: Pre-condition does not match!"
\end{lstlisting}

\section{Hoare logic}

In the previous section, we implemented assertions (\texttt{CAssert}) at the run-time (\texttt{eval}) level. The biggest disadvantage of that is we have to do a full evaluation to deduce some facts about programs; considering the assertion example of the \texttt{countToB} program, it has to actually evaluate the program to assert something. This motivates the need for an additional evaluation strategy that will allow us to deduce facts about programs without doing a full evaluation.

Some programming languages, like Python, don't have a compile step and the \texttt{eval} function we provided is kind of equivalent to evaluating programs in Python. But some programming languages do have a compile step, like C or Haskell, and this compilation step can be beneficial in that it can do additional different checks, e.g., type checks. That's what we'll do here - implement a "compile"-time check (just another evaluation strategy) using some of the rules in Hoare's logic, and this check can be used to check the validity of a program, before fully evaluating it.

We list the rules of Hoare logic, some of which are outlined in the original paper\cite{b8}.

\[
\AxiomC{}
\RightLabel{(H-Skip)}
\UnaryInfC{$\{P\}\texttt{skip}\{P\}$}
\DisplayProof
\quad
\AxiomC{}
\RightLabel{(H-Assign)}
\UnaryInfC{$\{P[E/x]\}x \texttt{ := } E\{P\}$}
\DisplayProof
\]
\hfill
\[
\AxiomC{$P_1 \to P_2 , \{P_2\}S\{Q_2\} , Q_2 \to Q_1$}
\RightLabel{(H-Consequence)}
\UnaryInfC{$\{P_1\} S \{Q_1\}$}
\DisplayProof
\quad
\AxiomC{$\{P\}S\{Q\} , \{Q\}T\{R\}$}
\RightLabel{(H-Sequence)}
\UnaryInfC{$\{P\}S;T\{R\}$}
\DisplayProof
\]
\hfill
\[
\AxiomC{$\{B \land P\}S\{Q\} , \{\neg B \land P\}T\{Q\}$}
\RightLabel{(H-Conditional)}
\UnaryInfC{$\{P\}\texttt{if }B\texttt{ then }S\texttt{ else } T\{Q\}$}
\DisplayProof
\quad
\quad
\AxiomC{$\{B \land P\}S\{P\}$}
\RightLabel{(H-While)}
\UnaryInfC{$\{P\}\texttt{while }B\texttt{ do }S\{\neg B \land P\}$}
\DisplayProof
\]

\subsection{Implementation}

We represent the Hoare triple as a product of \texttt{Command a}, and the two conditions as \texttt{PropCalc (FOL a)}. As before, triples should not be constructed using \texttt{HoareTriple}, rather through the rules that we provide next.

\begin{lstlisting}
data HoareTriple a = HoareTriple (PropCalc (FOL a)) (Command a) (PropCalc (FOL a))
type ESHT a = Either String (HoareTriple a)
\end{lstlisting}

\subsubsection{H-Skip}

The Haskell implementation of the Hoare skip rule can be represented as follows:

\begin{lstlisting}
hoareSkip :: PropCalc (FOL a) -> ESHT a
hoareSkip q = Right $ HoareTriple q CSkip q
\end{lstlisting}

Note that the rule accepts well-formed formulas instead of proven ones, as the validity of the precondition is not being checked; the triple merely states that assuming some precondition, a command produces some postcondition.

\begin{lstlisting}
> putStrLn $ pr $ hoareSkip (PropVar (Eq (Var A) (S (S (S Z)))))
{A=SSS0} ; {A=SSS0} ✓
\end{lstlisting}

\subsubsection{H-Assign}

Let $Q[E/V]$ denote the expression $Q$ in which each free occurrence of $V$ is replaced with $E$. Given an assignment command $V \texttt{ := } E$, it should produce the triple where the precondition is $Q[E/V]$ and the postcondition is $Q$.

\begin{lstlisting}
hoareAssignment :: Eq a => a -> Arith a -> PropCalc (FOL a) -> ESHT a
hoareAssignment v e q = Right $ HoareTriple (fromProof (substPropCalc (Proof q) (Var v) e)) (CAssign v e) q
\end{lstlisting}

$Q$ exactly corresponds to \texttt{substPropCalc} - the implementation relies on \texttt{PropCalc}. Example deduction:

\begin{lstlisting}
> putStrLn $ pr $ hoareAssignment A (Plus (Var B) (S Z)) (And (PropVar (Eq (Var A) (S (S Z)))) (PropVar (Eq Z Z)))
{B+S0=SS0∧0=0} A := B+S0; {A=SS0∧0=0} ✓
\end{lstlisting}

\subsubsection{H-Consequence}

The consequence rule can be used to strengthen (make more specific) a precondition and/or weaken (make more generic) a postcondition in a Hoare triple. The implication in the consequence rule represents an "evaluation" from the previously defined logic. In other words, the rule provides a way to transform a Hoare triple by embedding the result of an evaluation of a logic (another system) into the same Hoare triple.

\begin{lstlisting}
hoareConsequence :: Eq a => Proof (PropCalc (FOL a)) -> HoareTriple a -> Proof (PropCalc (FOL a)) -> ESHT a
hoareConsequence (Proof (Imp p1 p2)) (HoareTriple p2' c q2) (Proof (Imp q2' q1))
  | p2 == p2' && q2 == q2' = Right $ HoareTriple p1 c q1
hoareConsequence _ _ _ = Left "hoareConsequence: Cannot construct proof"
\end{lstlisting}

For example, considering the \texttt{hoareAssignment} example we saw earlier, we can weaken its postcondition:

\begin{lstlisting}
> let ht = hoareAssignment A (Plus (Var B) (S Z)) (And (PropVar (Eq (Var A) (S (S Z)))) (PropVar (Eq Z Z)))
{B+S0=SS0∧0=0} A := B+S0; {A=SS0∧0=0} ✓
> let pre = ruleFantasy (And (PropVar (Eq (Plus (Var B) (S Z)) (S (S Z)))) (PropVar (Eq Z Z))) return
⊢ B+S0=SS0∧0=0→B+S0=SS0∧0=0 ✓
> let post = ruleFantasy (And (PropVar (Eq (Var A) (S (S Z)))) (PropVar (Eq Z Z))) ruleSepL
⊢ A=SS0∧0=0→A=SS0 ✓
> putStrLn $ pr $ join $ hoareConsequence <$> pre <*> ht <*> post
{B+S0=SS0∧0=0} A := B+S0; {A=SS0} ✓
\end{lstlisting}

For another example, we can strengthen the precondition:

\begin{lstlisting}
> let ht = hoareAssignment A (Plus (Var B) (S Z)) (PropVar (Eq (Var A) (S (S Z))))
{B+S0=SS0} A := B+S0; {A=SS0} ✓
> let pre = ruleFantasy (And (PropVar (Eq (Plus (Var B) (S Z)) (S (S Z)))) (PropVar (Eq Z Z))) ruleSepL
⊢ B+S0=SS0∧0=0→B+S0=SS0 ✓
> let post = ruleFantasy (PropVar (Eq (Var A) (S (S Z)))) return
⊢ A=SS0→A=SS0 ✓
> putStrLn $ pr $ join $ hoareConsequence <$> pre <*> ht <*> post
{B+S0=SS0∧0=0} A := B+S0; {A=SS0} ✓
\end{lstlisting}

\subsubsection{H-Sequence}

For the Hoare sequence rule, given two Hoare triples, the postcondition of the first triple must be equivalent to the precondition of the second triple, for some definition of equivalent; in this specific case, we rely on Haskell's \texttt{Eq}.

\begin{lstlisting}
hoareSequence :: Eq a => HoareTriple a -> HoareTriple a -> ESHT a
hoareSequence (HoareTriple p c1 q1) (HoareTriple q2 c2 r)
  | q1 == q2  = Right $ HoareTriple p (CSequence c1 c2) r
hoareSequence _ _ = Left "hoareSequence: Cannot construct proof"
\end{lstlisting}

Several commands can be sequenced as follows:

\begin{lstlisting}
> let c1 = hoareAssignment B Z (And (PropVar $ Eq (Var B) Z) (PropVar $ Eq (Var A) (Var A)))
{0=0∧A=A} B := 0; {B=0∧A=A} ✓
> let c2 = hoareAssignment C (Var A) (And (PropVar $ Eq (Var B) Z) (PropVar $ Eq (Var C) (Var A)))
{B=0∧A=A} C := A; {B=0∧C=A} ✓
> putStrLn $ pr $ join $ hoareSequence <$> c1 <*> c2
{0=0∧A=A} B := 0; C := A; {B=0∧C=A} ✓
\end{lstlisting}

\subsubsection{H-Conditional}

The Hoare conditional rule can be implemented as follows:

\begin{lstlisting}
hoareConditional :: Eq a => HoareTriple a -> HoareTriple a -> ESHT a
hoareConditional (HoareTriple (And b1 p1) c1 q1) (HoareTriple (And (Not b2) p2) c2 q2)
  | b1 == b2 && p1 == p2 && q1 == q2  = Right $ HoareTriple p1 (CIfElse b1 c1 c2) q1
hoareConditional _ _ = Left "hoareConditional: Cannot construct proof"
\end{lstlisting}

For the purposes of example, we start by considering the command:

\begin{lstlisting}
> putStrLn $ pr $ CIfElse (Not (PropVar $ Eq (Var A) Z)) CSkip (CAssign A (Plus (Var A) (S Z)))
if (¬A=0) then {;} else {A := A+S0;};
\end{lstlisting}

Following the rule to construct a Hoare triple for this command, we can notice that the conditionals $B$ and $P$ are given in the code, so we have to construct the two triples $\{A \neq 0 \land P\} \texttt{ ; } \{Q\}$ and $\{\neg(A \neq 0) \land P\} A \texttt{ := } 1 + A; \{Q\}$.

\begin{lstlisting}
> let ht1 = hoareSkip (And (Not $ PropVar (Eq (Var A) Z)) (PropVar $ Eq Z Z))
{¬A=0∧0=0} ; {¬A=0∧0=0} ✓
> let ht2 = hoareAssignment A (S (Var A)) (And (Not $ PropVar (Eq (Var A) Z)) (PropVar $ Eq Z Z))
{¬SA=0∧0=0} A := SA; {¬A=0∧0=0} ✓
\end{lstlisting}

The next thing to do is to transform the precondition of the second triple to match the first one.

\begin{lstlisting}
> let prf1 = ruleFantasy (And (PropVar (Eq (Var A) Z)) (PropVar $ Eq Z Z)) (\pq -> join $ ruleJoin <$> (axiom1 (Var A) >>= ruleSpec (Var A)) <*> ruleSepR pq)
⊢ A=0∧0=0→¬SA=0∧0=0 ✓
> let prf2 = ruleFantasy (And (Not $ PropVar (Eq (Var A) Z)) (PropVar $ Eq Z Z)) return
⊢ ¬A=0∧0=0→¬A=0∧0=0 ✓
\end{lstlisting}

Given these two proofs, we can now transform the triple easily, and conclude our proof:

\begin{lstlisting}
> let ht3 = join $ hoareConsequence <$> prf1 <*> ht2 <*> prf2
{A=0∧0=0} A := SA; {¬A=0∧0=0} ✓
> putStrLn $ pr $ join $ hoareConditional <$> ht3 <*> ht1
{0=0} if (A=0) then {A := SA;} else {;}; {¬A=0∧0=0} ✓
\end{lstlisting}

\subsubsection{H-While}

In this part, we provide the implementation of the last, most important, and complex rule.

\begin{lstlisting}
hoareWhile :: Eq a => HoareTriple a -> ESHT a
hoareWhile (HoareTriple (And b p1) c p2)
  | p1 == p2  = Right $ HoareTriple p1 (CWhile b c) (And (Not b) p1)
hoareWhile _ = Left "hoareWhile: Cannot construct proof"
\end{lstlisting}

The implementation for constructing the rule is straightforward, but the tricky part is that we have to properly construct the precondition and the postcondition so that they match the rule.

\pagebreak

Again, for the purposes of example, consider the following command:

\begin{lstlisting}
> putStrLn $ pr (CWhile (PropVar $ Eq Z Z) CSkip :: Command Integer)
while (0=0) do {;};
\end{lstlisting}

Similarly, as we did in the example for the H-Conditional rule, following the rule to construct a Hoare triple for this command, we get that $B$ is \texttt{(PropVar \$ Eq Z Z)} and $C$ is \texttt{CSkip}.

\begin{lstlisting}
> let ht1 = hoareSkip (PropVar $ Eq Z Z) :: ESHT Integer
{0=0} ; {0=0} ✓
> let pre = ruleFantasy (And (PropVar $ Eq Z Z) (PropVar $ Eq Z Z)) ruleSepR :: ESP (FOL Integer)
⊢ 0=0∧0=0→0=0 ✓
> let post = ruleFantasy (PropVar $ Eq Z Z) return :: ESP (FOL Integer)
⊢ 0=0→0=0 ✓
> let ht2 = join $ hoareConsequence <$> pre <*> ht1 <*> post
{0=0∧0=0} ; {0=0} ✓
\end{lstlisting}

Now we can deduce a triple which states that from an infinite loop, anything follows (contradiction); our implementation does not support proving total correctness (termination).

\begin{lstlisting}
> putStrLn $ pr $ ht2 >>= hoareWhile
{0=0} while (0=0) do {;}; {¬0=0∧0=0} ✓
\end{lstlisting}

\subsection{A complete proof example}

We provide a proof for the program \texttt{countToB}, namely that $B$ will be equal to $A$ after its execution. According to the program definition, we need to apply the rules in order: H-Sequence, H-Assign, H-While. Looking at the while rule, we can determine the values of $B$ and $S$ based on the command; meanwhile, one valid invariant would be $\exists C: A + C = B$.

\[
\AxiomC{$\{\neg(A = B) \land \exists C: A + C = B\}A\texttt{ := }S(A) \ \{\exists C: A + C = B\}$}
\RightLabel{(proof)}
\UnaryInfC{$\{P\}\texttt{while } \neg(A = B) \texttt{ do }A\texttt{ := }S(A) \ \{\neg (\neg (A = B)) \land \exists C: A + C = B\}$}
\DisplayProof
\]

Further, since this rule has to be followed in sequence with the assignment rule, it dictates the conditions for it:

\[
\AxiomC{}
\RightLabel{(step2)}
\UnaryInfC{$\{\exists C: 0 + C = B\} A \texttt{ := } 0 \{\exists C: A + C = B\}$}
\DisplayProof
\]

The only thing remaining to prove is the hypothesis for the H-While rule, so we have the following code:

\begin{lstlisting}
proof = join $ hoareSequence <$> step2 <*> step3
step2 = hoareAssignment A Z (PropVar (Exists C (PropVar $ Eq (Plus (Var A) (Var C)) (Var B))))
step3 = join $ hoareWhile <$> step4
\end{lstlisting}

To start figuring out \texttt{step4}, we consider the following code:

\begin{lstlisting}
> let step5 = hoareAssignment A (S (Var A)) (PropVar (Exists C (PropVar $ Eq (Plus (Var A) (Var C)) (Var B))))
{∃C:(SA+C=B)} A := SA; {∃C:(A+C=B)} ✓
\end{lstlisting}

The resulting triple is close but does not exactly match the hypothesis of \texttt{proof}'s hypothesis, more specifically the precondition. We apply the H-Consequence rule, and to do that, we need to provide a proof for the statement:
$$\neg (A = B) \land \exists C: A + C = B \to \exists C: S(A) + C = B$$

\begin{lstlisting}
pre   = ???
post  = ruleFantasy (PropVar (Exists C (PropVar $ Eq (Plus (Var A) (Var C)) (Var B)))) return
step4 = join $ hoareConsequence <$> pre <*> step5 <*> post
\end{lstlisting}

Here's the complete proof for the preconditional part, relying on the previously derived \texttt{theorem}:

\begin{lstlisting}
pre = ruleFantasy (And (Not (PropVar $ Eq (Var A) (Var B))) (PropVar (Exists C (PropVar $ Eq (Plus (Var A) (Var C)) (Var B))))) $ \premise -> do
    -- ⊢ SA+C=A+SC
    eq1 <- theorem >>= ruleSpec (Var C) >>= ruleSpec (Var A) >>= ruleSymmetry
    -- ⊢ ¬¬∃C:(A+C=B)
    step1 <- ruleSepR premise >>= ruleDoubleTildeIntro
    -- ⊢ ¬¬A+SC=B
    step2 <- applyFOLRule [GoLeft] (\x -> ruleInterchangeR x >>= ruleSpec (S (Var C))) [] step1
    -- ⊢ A+SC=B
    eq2 <- ruleDoubleTildeElim step2
    -- ⊢ SA+C=B
    eq3 <- ruleTransitivity eq1 eq2
    -- ⊢ ¬A=B∧∃C:(A+C=B)→∃C:(SA+C)=B
    ruleExistence C [] eq3
\end{lstlisting}

Finally, \texttt{proof} produces the triple:
$$\{\exists C: 0+C=B\} A \texttt{ := } 0 \texttt{; while } \neg(A = B) \texttt{ do }A\texttt{ := }S(A) \ \{\neg (\neg (A = B)) \land \exists C: A + C = B\}$$

\section{Conclusion and future work}

Compile-time, run-time, etc. are all about having evaluations at different levels. There is still computation going on, but the computation strategies are different at various computation levels. A full evaluation of \texttt{Command} can be expensive, and sometimes even not terminate, and we wanted a way to deduce propositions without doing a full evaluation. We achieved this by providing an implementation of Hoare logic.

The biggest disadvantage of the provided implementation is the manual construction of proofs, which is a bit tedious as it feels like working with machine code (one similar system is Metamath\cite{b9}). Ideally, we would rely on an automated theorem prover that would derive the necessary formulas. Further, the underlying number-theoretical system uses Peano's axioms which is limited to natural numbers - we could extend it to integers for better expressibility. In addition, we could increase the expressiveness of the logical system by allowing user definitions to be added. Finally, we could add more commands to the imperative language and correspondingly, rules to the Hoare logic.

Even though the mathematical formulas for these systems look simple, implementing them is a different matter. Looking at the implementation, we see how complex the implementation details of these systems can get. Nevertheless, this introductory paper may serve as a good starting point toward building systems of similar nature.

\section{Conflict of interest}

The author declares that they have no conflict of interest.

\end{document}